\documentclass{ws-procs9x6}
\newcommand{\cal}{\mathcal}

\begin{document}

\title{Quantum Integrable System with Multi-components in Two-Dimension \footnote{\uppercase{T}his work is
partially supported by \uppercase{NSF} of \uppercase{C}hina
90103002.}}

\author{Mu-Lin Yan}

\address{Center for Fundamental Physics, \\
University of Science and Technology of China, \\
Hefei, Anhui 230026, China\\
E-mail: mlyan@ustc.edu.cn}

\author{Bao-Heng Zhao}

\address{Department of Physics,\\
Graduate School of the Chinese Academy of Science,\\
 Beijing 100039, China}


\maketitle

\abstracts{ A quantum N-body problem with 2-component in
(2+1)-dimension deduced from integrable model in (2+1) dimension
is investigated. The Davey-Stewartson 1(DS1) system[Proc. R. Soc.
London, Ser. A {\bf 338}, 101 (1974)] is an integrable model in
two dimensions. A quantum DS1 system with 2 colour-components in
two dimensions has been formulated. This two-dimensional problem
has been reduced to two one-dimensional many-body problems with 2
colour-components. The solutions of the two-dimensional problem
under consideration has been constructed from the resulting
problems in one dimensions. For latters with the $\delta
$-function interactions and being solved by the Bethe-Yang ansatz,
we introduce symmetrical and antisymmetrical Young operators of
the permutation group  and obtain the exact solutions for the
quantum DS1 system. The application of the solusions is
discussed.}

\section{Introduction}
It is well known that the high-dimensional quantum integrability
of some one body system is very meaningful for precisely
understanding physics. Say, one of the typical examples is to
fined out the electron wave-functions in the hydrogen atom: there
is one body in the atom, i.e., electron, and its motion equation
is (3+1) dimensional Schrodinger equation . The key step to solve
this problem is to use  variable separate ansatz for the one-body
wave-function, $\Psi (r,\theta , \phi )=R(r)P(\theta ) \Phi (\phi
)$ to reduce the 3-Dimensional (3-D) one-body problem to
1-Dimensional(1-D) one-body problem. It is interesting and
remarkable that the one body variable separate ansatz method can
be extend to the $N$-body case. Namely, for some special high
dimensional quantum N-body models (say the Davey-Stewartson 1
sestem in 2-D), they could be reduced to 1-D problems in terms to
use the ansatz. In this paper, we try to further extend this sort
of ansatz to multi-component case, and to solve it by suing
Bethe-Yang ansatz.

  The Davey-Stewartson 1 (DS1) system is an integrable model in
space of two spatial and one temporal dimensions ((2+1)D). The
quantized DS1 system with scalar fields (1 component or shortly
1C) can be formulated in  terms of the Hamiltonian of quantum
many-body problem in two dimensions, and some of them can be
solved exactly\cite{pang,yan}. Particularly, it has been shown in
ref.\cite{yan} that these 2D quantum $N$-body system with
1C-fields can be reduced to the solvable one-dimensional (1D)
quantum $N$-body systems with 1C-fields and with two-body
potentials\cite{ols}. Thus through solving 1D quantum $N$-body
problems with 1C-fields we can get the solutions for 2D quantum
$N$-body problems with 1C-fields. Here the key step is to separate
the spatial variables of 2D quantum $N$-body problems with
1C-fields by constructing an ansatz\cite{pang,yan}
$$
\Psi (\xi_1, \cdots , \xi_N,  \eta_1, \cdots , \eta_N) =
\prod_{i<j} (1-\frac{c}{4}\epsilon (\xi_{ij})\epsilon (\eta_{ij}))
  X(\xi_1, \cdots , \xi_N)Y(\eta_1, \cdots, \eta_N)
$$
where $\xi_{ij}=\xi_i-\xi_j$ and $\eta_{ij}=\eta_i-\eta_j$. This
ansatz will be called the N-body variable-separation ansatz.  The
N-body variable-separation ansatz can be thought of as the
extension of 1-body variable-separation ansatz. Since the N-body
problems are much more complicated than the 1-body problems, it
will be highlly nontrivial to construct a N-body
variable-separation ansatz. Ref.\cite{pang} provided the first
example for it and showed that the idea of variable- separating
works indeed for the N-body problems  induced from the DS1  system
. \par In this paper, we intend to generalize the above idea to
multi-components DS1 system, namely to construct a new N-body
variable-separation ansatz for the  multi-components case  and to
solve a specific model  of 2D quantum DS1 system with
multi-components. \par 1D N-body model with  2-components has been
investigated for long time\cite{yang}\cite{cal}. The most famous
one is the model with delta-function interaction between
2C-fermions\cite{yang}. It was solved by the Bethe-Yang
ansatz\cite{bethe,yang}(or nested Bethe ansatz) and leads to the
Yang-Baxter equation and its thermodynamics studies\cite{lai,sch}
because of the completeness of the Bethe ansatz solutions. In this
paper, for definiteness, we shall study specific 2D quantum
$N$-body system with 2C-fields associated with the DS1 system.
This quantum $N$-body problem under consideration can be reduced
to two 1D quantum $N$-body problems with 2C-fields of
ref.\cite{yang} and then be exactly solved by using an appropriate
N-body variable-separation ansatz and the Bethe-Yang ansatz. \par

\section{ Quantization of DS1 System with Two Components in 2-Dimension}

Following usual DS1 equation\cite{pang,ds}, the equation for the
DS1 system with two components reads
\begin{equation}
i{\bf \dot{q}} =-\frac{1}{2} (\partial _{x}^{2}
+\partial_{y}^{2}){\bf q}+
 iA_{1}{\bf q}+iA_{2}{\bf q},
\end{equation}
where ${\bf q}$ has two colour components,
\begin{equation}
{\bf q}=
    \left( \begin{array}{c}
    q_{1}\\
    q_{2}\\
    \end{array} \right),
\end{equation}
and
\begin{eqnarray*}
(\partial _{x}-\partial
_{y})A_{1}&=&-ic(\partial_{x}+\partial_{y})({\bf q^{\dag}q})\\
(\partial _{x}+\partial
_{y})A_{2}&=&ic(\partial_{x}-\partial_{y})({\bf q^{\dag}q})
\end{eqnarray*}
where notation ${\bf \dag }$ means the hermitian transposition,
and $c$ is the coupling constant. Introducing the coordinates $\xi
=x+y, \eta =x-y,$ we have
\begin{eqnarray}
A_{1}&=&-ic\partial _{\xi}\partial _{\eta}^{-1}({\bf q^{\dag}q})
        -iu_{1}(\xi)  \\
A_{2}&=&ic\partial _{\eta}\partial _{\xi}^{-1}({\bf q^{\dag}q})
        +iu_{2}(\eta)
\end{eqnarray}
where
\begin{equation}
\partial_{\eta}^{-1}({\bf
q^{\dag}q})=\frac{1}{2}(\int_{-\infty}^{\eta} d\eta^{\prime}-
\int^{\infty}_{\eta} d\eta^{\prime}){\bf
q^{\dag}}(\xi,\eta^{\prime},t) {\bf q}(\xi,\eta^{\prime},t),
\end{equation}
and $u_{1}$ and $u_{2}$ are constants of integration. According to
ref.\cite{yan}, we choose them as
\begin{eqnarray}
u_{1}(\xi)&=&\frac{1}{2}\int
d\xi^{\prime}d\eta^{\prime}U_{1}(\xi-\xi^{\prime})
           {\bf q^{\dag}}(\xi^{\prime},\eta^{\prime},t)
               {\bf q}(\xi^{\prime},\eta^{\prime},t)  \\
u_{2}(\eta)&=&\frac{1}{2}\int
d\xi^{\prime}d\eta^{\prime}U_{2}(\eta-\eta^{\prime})
           {\bf q^{\dag}}(\xi^{\prime},\eta^{\prime},t)
               {\bf q}(\xi^{\prime},\eta^{\prime},t).
\end{eqnarray}
Thus eq.(1) can be written as
\begin{eqnarray}
i{\bf \dot{q}}=-(\partial_{\xi}^{2}+\partial_{\eta}^{2})
  {\bf q}+c[\partial_{\xi}\partial_{\eta}^{-1}({\bf q^{\dag}q})
+\partial_{\eta}\partial_{\xi}^{-1}({\bf q^{\dag}q})]{\bf q}
\nonumber \\
+\frac{1}{2}\int
d\xi^{\prime}d\eta^{\prime}[U_{1}(\xi-\xi^{\prime})+
+U_{2}(\eta-\eta^{\prime})]({\bf q}^{\dag \prime}{\bf
q}^{\prime}){\bf q},
\end{eqnarray}
where ${\bf q}^{\prime}={\bf q}(\xi^{\prime},\eta^{\prime},t)$. We
quantize the system with the canonical commutation relations
\begin{equation}
 [ q_{a}(\xi,\eta,t), q^{\dag}_{b}(\xi^{\prime},
\eta^{\prime},t)]_{\pm}
= 2\delta_{_{ab}}\delta(\xi-\xi^{\prime})\delta(\eta-\eta^{\prime}), \\
\end{equation}
\begin{equation}
 [ q_{a}(\xi,\eta,t), q_{b}(\xi^{\prime},\eta^{\prime},t)]_{\pm}
=0.
\end{equation}
where $a,b = 1$ or $2$, $[ , ]_{+}$ and $[ , ]_{-}$ are
anticommutator
 and commutator respectively.
Then eq.(8) can be written in the form
\begin{equation}
\dot{{\bf q}}=i[H,{\bf q}]
\end{equation}
where $H$ is the Hamiltonian of the system
\begin{eqnarray}
H=\frac{1}{2}\int d\xi d\eta
\begin{array}{c}${\LARGE(}$\end{array}
 -{\bf q}^{\dag}
(\partial_{\xi}^{2}+\partial_{\eta}^{2})
  {\bf q}+\frac{c}{2}{\bf q}^{\dag}[(\partial_{\xi}\partial_{\eta}^{-1}
  +\partial_{\eta}\partial_{\xi}^{-1})({\bf q^{\dag}q})]{\bf q}
\nonumber \\
+\frac{1}{4}\int d\xi^{\prime}d\eta^{\prime}{\bf q}^{\dag}
[U_{1}(\xi-\xi^{\prime}) +U_{2}(\eta-\eta^{\prime})]({\bf
q}^{\prime \dag}{\bf q}^{\prime}){\bf
q}\begin{array}{c}${\LARGE)}$\end{array}.
\end{eqnarray}
The $N$-particle eigenvalue problem is
\begin{equation}
H\mid \Psi \rangle =E\mid \Psi \rangle
\end{equation}
where
\begin{eqnarray}
\hspace{-0.5cm}\mid \Psi \rangle&=&\int d\xi_{1}d\eta_{1}\ldots
d\xi_{N}d\eta_{N}
\nonumber \\
&&\times \sum_{a_{1}\ldots a_{N}} \Psi_{a_{1}\ldots
a_{N}}(\xi_{1}\eta_{1} \ldots \xi_{N}\eta_{N})
{\bf q}^{\dag}_{a_{1}}(\xi_{1}\eta_{1})\ldots {\bf
q}^{\dag}_{a_{N}}(\xi_{N}\eta_{N}) \mid 0 \rangle.
\end{eqnarray}
The $N$-particle wave function $ \Psi_{a_{1}\ldots a_{N}}$ is
defined by eq.(14), which satisfies the $N$-body Schr\"{o}dinger
equation
\begin{eqnarray}
-\sum_{i}(\partial_{\xi_{i}}^{2}+\partial_{\eta_{i}}^{2})
 \Psi_{a_{1}\ldots a_{N}}+c\sum_{i<j}
 [\epsilon (\xi_{ij})\delta^{\prime}(\eta_{ij})
  +\epsilon (\eta_{ij})\delta^{\prime}(\xi_{ij})] \Psi_{a_{1}\ldots a_{N}}
  \nonumber  \\
  +\sum_{i<j}[U_{1}(\xi_{ij})+U_{2}(\eta_{ij})] \Psi_{a_{1}\ldots a_{N}}
  =E \Psi_{a_{1}\ldots a_{N}}
\end{eqnarray}
where $\xi_{ij}=\xi_{i}-\xi_{j},
\delta^{\prime}(\xi_{ij})=\partial_{\xi_{i}} \delta (\xi_{ij})$,
and $\epsilon (\xi_{ij})=1$ for $\xi_{ij}>0, 0$ for $\xi_{ij}=0,
-1$ for $\xi_{ij}<0.$ Since there are products of distributions in
eq.(15), an appropriate regularezation for avoiding uncertainty is
necessary. This issue has been discussed in ref.\cite{zhao}.

\section{Variable Separation of Quantum DS1 with Two Components
and Bethe-Yang Ansatz}

Our purpose is to solve the $N$-body Schr\"{o}dinger equation
(15). The results in ref.\cite{yan} remind us that we can make the
following ansatz
\begin{eqnarray}
\Psi_{a_{1}\ldots a_{N}} &=&
\sum_{ a_{1}^{\prime} \ldots a_{N}^{\prime} \atop
       b_{1}^{\prime} \ldots b_{N}^{\prime}}
 \prod_{i<j} (1-\frac{c}{4}\epsilon (\xi_{ij})\epsilon (\eta_{ij}))
{\cal M}_{a_{1}\ldots a_{N}, a_{1}^{\prime}\ldots a_{N}^{\prime}}
 {\cal N}_{a_{1}\ldots a_{N}, b_{1}^{\prime}\ldots b_{N}^{\prime}} \nonumber  \\
&&\times X_{ a_{1}^{\prime}\ldots a_{N}^{\prime}}(\xi_{1}\ldots
\xi_{N}) Y_{ b_{1}^{\prime}\ldots b_{N}^{\prime}}(\eta_{1}\ldots
\eta_{N})
\end{eqnarray}
where ${\cal M}$ and ${\cal N}$ are matrices being independent of
$\xi$ and $\eta$, and both $X_{ a_{1}\ldots a_{N}}(\xi_{1}\ldots
\xi_{N})$ and $Y_{ b_{1}\ldots b_{N}}(\eta_{1}\ldots \eta_{N})$
are one-dimensional wave functions of N-bodies. Substituting
eq.(16) into eq.(15), we abtian
\begin{eqnarray}
-\sum_{i} \partial_{\xi_{i}}^{2} X_{a_{1}\ldots a_{N}} +\sum_{i<j}
U_{1}(\xi_{ij}) X_{a_{1}\ldots a_{N}}&=&E_{1} X_{a_{1}\ldots
a_{N}}
\\
-\sum_{i} \partial_{\eta_{i}}^{2} Y_{b_{1}\ldots b_{N}}
+\sum_{i<j} U_{2}(\eta_{ij}) Y_{b_{1}\ldots
b_{N}}&=&E_{2}Y_{b_{1}\ldots b_{N}}
\end{eqnarray}
where $U_{1}(\xi_{ij})$ and $U_{2}(\eta_{ij})$ are two-body
potentials, eqs. (17) (18) are one-dimensional $N$-body
Schr\"{o}dinger equations and $E_{1}+E_{2}=E$. Above derivation
indicates that the two-dimensional $N$-body Schr\"{o}dinger
equation (15) has been reduced into two one-dimentional
 $N$-body Schr\"{o}dinger equations. Namely, the variables in the
two-dimentional $N$-body wave function $\Psi_{a_{1}\ldots a_{N}}$
have been separated.

At this stage ${\cal M}$ and ${\cal N}$ are unknown temporarily.
 It is expected that for any given pair of exactly solvable 1D N-body
 problems and the  correspondent solutions,
we could construct the solutions $\Psi _{A_{1} \ldots A_{N}}$ for
2D N-body problems eq.(15) through constructing an appropriate
${\cal M} \times {\cal N}$-matrix.
It has been known that the 1D  N-body problem in the form of (17)
or (18) can be solved exactly for a class of
potentials\cite{yang,cal,sut}.  To illustrate the construction of
${\cal M} \times {\cal N}$-matrix, we  take  both potentials in
(17) and (18) the delta functions $U_{1}(\xi_{ij})=2g\delta
(\xi_{ij})$ and $U_{2}(\eta_{ij})=2g\delta (\eta_{ij})$ ($g>0$,
the coupling constant). Then eqs.(17) and (18) become
\begin{eqnarray}
-\sum_{i} \partial_{\xi_{i}}^{2} X_{a_{1}\ldots a_{N}} +2g
\sum_{i<j} \delta(\xi_{ij}) X_{a_{1}\ldots a_{N}} &=&E_{1}
X_{a_{1}\ldots a_{N}}
\\
-\sum_{i} \partial_{\eta_{i}}^{2} Y_{b_{1}\ldots b_{N}} +2g
\sum_{i<j} \delta(\eta_{ij}) Y_{b_{1}\ldots b_{N}}
&=&E_{2}Y_{b_{1}\ldots b_{N}}.
\end{eqnarray}
As $X$ and $Y$ are wave functions of Fermions with two components,
denoted by $X^{F}$ and $Y^{F}$, the problem has been solved by
Yang long ago\cite{yang} (more explicitly, see ref.\cite{gu} and
ref.\cite{fan}). According to the Bethe-Yang ansatz, the continual
solution of eq.(9) in the region of
$0<\xi_{Q_{1}}<\xi_{Q_{2}}<\ldots <\xi_{Q_{N}}<L$ reads
\begin{eqnarray}
X^{F}&=&\sum_{P}
\alpha_{P}^{(Q)}\exp\{i[k_{P_{1}}\xi_{Q_{1}}+\ldots+k_{P_{N}}\xi_{Q_{N}}
]\} \nonumber  \\
&=&\alpha_{_{12\ldots
N}}^{(Q)}e^{i(k_{1}\xi_{Q_{1}}+k_{2}\xi_{Q_{2}}
+\ldots+k_{N}\xi_{Q_{N}})} +\alpha_{_{21\ldots
N}}^{(Q)}e^{i(k_{2}\xi_{Q_{1}}+k_{1}\xi_{Q_{2}}
+\ldots+k_{N}\xi_{Q_{N}})}
\nonumber  \\
& &+(N!-2) \begin{array}{c} $others terms$ \end{array}
\end{eqnarray}
where  $X^{F}\in \{ X^{F}_{a_{1}\ldots a_{N}} \}$,
$P=[P_{1},P_{2},\cdots,P_{N}]$ and $Q=[Q_{1},Q_{2},\cdots,Q_{N}]$
are two permutations of the integers $1,2,\ldots ,N$, and
\begin{eqnarray}
\alpha^{(Q)}_{\ldots ij \ldots}&=&Y^{lm}_{ji} \alpha^{(Q)}_{\ldots
ji \ldots},
  \\
Y^{lm}_{ji}&=&\frac{-i(k_{j}-k_{i})P^{lm}+g}{i(k_{j}-k_{i})-g}
\end{eqnarray}
The eigenvalue is given by
\begin{equation}
E_{1}=k_{1}^{2}+k_{2}^{2}+\ldots +k_{N}^{2},
\end{equation}
where $\{ k_{i} \}$ are determined by the Bethe ansatz equations,
\begin{eqnarray}
& &e^{ik_{j}L}=\prod_{\beta =1}^{M}
\frac{i(k_{j}-\Lambda_{\beta})-g/2}
{i(k_{j}-\Lambda_{\beta})+g/2}   \\
& &\prod_{j=1}^{N} \frac{i(k_{j}-\Lambda_{\alpha})-g/2}
{i(k_{j}-\Lambda_{\alpha})+g/2} =-\prod_{\beta =1}^{M}
\frac{i(\Lambda_{\alpha}-\Lambda_{\beta})+g}
{i(\Lambda_{\alpha}-\Lambda_{\beta})-g}
\end{eqnarray}
with $\alpha =1,\ldots, M, j=1,\ldots, N.$ Through exactly same
procedures we can get the solution $Y^{F}$ and $E_{2}$ to eq.(20).

    As $X$ and $Y$ are Boson's wave-functions,
denoted by $X^{B}$ and $Y^{B}$, it is easy to be shown that
\begin{eqnarray}
X^{B}&=&\sum_{P}
\beta_{P}^{(Q)}\exp\{i[k_{P_{1}}\xi_{Q_{1}}+\ldots+k_{P_{N}}\xi_{Q_{N}}
]\}   \\
\beta^{(Q)}_{\ldots ij \ldots}&=&Z^{lm}_{ji} \beta^{(Q)}_{\ldots
ji \ldots},
  \\
Z^{lm}_{ji}&=&\frac{i(k_{j}-k_{i})P^{lm}+g}{i(k_{j}-k_{i})-g}
\end{eqnarray}
and the Bethe ansatz equations are as following\cite{fan}
\begin{eqnarray}
& &e^{ik_{j}L}=(-1)^{N+1}\prod_{i=1}^{N}
\frac{k_{j}-k_{i}+ig}{k_{j}-k_{i}-ig} \prod_{\beta =1}^{M}
\frac{\Lambda_{\beta}-k_{j}+ig/2}
{\Lambda_{\beta}-k_{j}-ig/2}   \\
& &\prod_{\alpha =1}^{M}
\frac{\Lambda_{\beta}-\Lambda_{\alpha}+ig}
{\Lambda_{\beta}-\Lambda_{\alpha}-ig} =(-1)^{M+1}\prod_{j =1}^{N}
\frac{\Lambda_{\beta}-k_{j}+ig/2} {\Lambda_{\beta}-k_{j}-ig/2}.
\end{eqnarray}
$Y^{B}$ is same as $X^{B}$. It is well known that $X^{F}$ and
$Y^{F}$($X^{B}$ and $Y^{B}$) are antisymmetrical (symmetrecal) as
the coordinates and the colour-indeces of the particles
interchanges each other simultaneously, instead of the coordinates
interchanges each other merely.

\section{Young Operator of Permutation Group}
 For permutation group $S_{N}: \{ e_{i}, i=1,\cdots,
N!\}$, the totally symmetrical Young operator is
\begin{equation}
{\cal O}_{N}=\sum_{i=1}^{N!} e_{i},
\end{equation}
and the totally antisymmetrical Young operator is
\begin{equation}
{\cal A}_{N}=\sum_{i=1}^{N!} (-1)^{P_{i}}e_{i}.
\end{equation}
The Young diagram for ${\cal O}_{N}$ is
{
\begin{tabular}{|l|l|l|l|l|r|}  \hline
 1  & 2 &3 &$\cdots$ & N  \\  \hline
\end{tabular}},
and for ${\cal A}_{N}$, it is
{\small
\begin{tabular}{|l|r|}  \hline
 1  \\  \hline
 2  \\  \hline
 \vdots  \\  \hline
 N  \\   \hline
\end{tabular}}.
To $S_{3}$, for example, we have
\begin{equation}
{\cal O}_{3}=1+P^{12}+P^{13}+P^{23}+P^{12}P^{23}+P^{23}P^{12}.
\end{equation}
\begin{equation}
{\cal A}_{3}=1-P^{12}-P^{13}-P^{23}+P^{12}P^{23}+P^{23}P^{12}.
\end{equation}
Lemma 1: $({\cal O}_{N}X_{F})(\xi_{1},\xi_{2},\cdots,\xi_{N})$ is
antisymmetrical with respect to the coordinate's interchanges of
$(\xi_{i} \longleftrightarrow \xi_{j})$.

\noindent Proof: From the definition of ${\cal O}_{N}$ (eq.(32)),
we have
\begin{equation}
{\cal O}_{N}P^{ab}=P^{ab}{\cal O}_{N}={\cal O}_{N}.
\end{equation}
To $N=3$ case, for example, the direct calculations show ${\cal
O}_{3}P^{12}=P^{12}{\cal O}={\cal O}_{3}, {\cal
O}_{3}P^{23}=P^{23}{\cal O}={\cal O}_{3}$ and so on. Using
eqs.(36) and (23), we have
\begin{equation}
{\cal O}_{N}Y^{lm}_{ij}=(-1){\cal O}_{N}.
\end{equation}
From eqs.(21) and (23), $X^{F}$ can be written as
\begin{eqnarray}
X^{F}&=&\{ e^{i(k_{1}\xi_{Q_{1}}+k_{2}\xi_{Q_{2}}
+\ldots+k_{N}\xi_{Q_{N}})}
+Y^{12}_{12}e^{i(k_{2}\xi_{Q_{1}}+k_{1}\xi_{Q_{2}}
+\ldots+k_{N}\xi_{Q_{N}})} \nonumber  \\
& &+Y^{23}_{13}Y^{12}_{12}e^{i(k_{2}\xi_{Q_{1}}
+k_{3}\xi_{Q_{2}}+k_{1}\xi_{Q_{3}} +\ldots+k_{N}\xi_{Q_{N}})} \nonumber\\
&& +(N!-3) \begin{array}{c} $other terms$ \end{array} \}
\alpha_{_{12\ldots N}}^{(Q)}
\end{eqnarray}
Using eqs.(37) and (38), we obtain
\begin{eqnarray}
&&({\cal O}_{N}X^{F})(\xi_{1},\cdots,\xi_{N})= \{
e^{i(k_{1}\xi_{Q_{1}}+k_{2}\xi_{Q_{2}} +\ldots+k_{N}\xi_{Q_{N}})} \nonumber\\
&&\quad -e^{i(k_{2}\xi_{Q_{1}}+k_{1}\xi_{Q_{2}} +\ldots+k_{N}\xi_{Q_{N}})}
+e^{i(k_{2}\xi_{Q_{1}} +k_{3}\xi_{Q_{2}}+k_{1}\xi_{Q_{3}}
+\ldots+k_{N}\xi_{Q_{N}})} \nonumber  \\
&&\quad+(N!-3)\, \, \mbox{other terms} \} {\cal O}_{N} \alpha_{_{12\ldots N}}^{(Q)}
\nonumber   \\
&&\quad =\sum_{P} (-1)^{P}
\exp\{i[k_{P_{1}}\xi_{Q_{1}}+\ldots+k_{P_{N}}\xi_{Q_{N}} ]\}({\cal
O}_{N} \alpha_{_{12\ldots N}}^{(Q)}).
\end{eqnarray}
Therefore we conclude that $({\cal
O}_{N}X^{F})(\xi_{1},\cdots,\xi_{N})$ is antisymmetrical with
respect to $(\xi_{i}\longleftrightarrow \xi_{j})$.

\noindent Lemma 2: $({\cal
A}_{N}X^{B})(\xi_{1},\xi_{2},\cdots,\xi_{N})$ is antisymmetrical
with respect to the coordinate's interchanges of $(\xi_{i}
\longleftrightarrow \xi_{j})$.

\noindent  Proof: Noting (see eqs.(33) (29) (27))
\begin{eqnarray}
{\cal A}_{N} P^{ab}=P^{ab}{\cal A}=-{\cal A}_{N},  \\
{\cal A}_{N} Z^{lm}_{ij}=(-1){\cal A}_{N},
\end{eqnarray}
we then have
\begin{eqnarray}
({\cal A}_{N}X^{B})(\xi_{1},\cdots,\xi_{N}) &=& \sum_{P} (-1)^{P}
\exp\{i[k_{P_{1}}\xi_{Q_{1}}+\ldots+k_{P_{N}}\xi_{Q_{N}} ]\} \nonumber\\
&&\quad \times ({\cal A}_{N} \beta_{_{12\ldots N}}^{(Q)}).
\end{eqnarray}
Then the Lemma is proved.

\section{The Solutions of the Problem}

 The ansatz of eq.(16) can be compactly written as
\begin{equation}
\Psi
= \prod_{i<j} (1-\frac{c}{4}\epsilon (\xi_{ij})\epsilon (\eta_{ij}))
  ({\cal M}X)  ({\cal N}Y)
\end{equation}
where $({\cal M}X)$ and $({\cal N}Y)$ are required to be
antisymmetrical under the interchanges of the coodinate vairables.
According to Lemmas 1 and 2, we see that
\begin{equation}
 {\cal M},{\cal N} =\left\{
\begin{array}{ccc}{\cal O}_{N} & & \, \, \, \, $for 1D Fermion$  \\
                  {\cal A}_{N} & & $for 1D Boson.$
\end{array} \right.
\end{equation}
As the DS1 fields $q_{a}(\xi \eta)$ in eq.(1) are (2+1)D Bose fields,
the commutators ($[ , ]_{-}$, see (9) and (10)) are used to quantized
the system and the 2D N-body wave functions denoted in $\Psi ^{B}$
must be symmetrical under the colour-interchang $(a_{i} \longleftrightarrow
a_{j})$ and the coordinate-interchange $((\xi_{i} \eta_{i})
\longleftrightarrow  (\xi_{j} \eta_{i}))$. Namly, the 2D Bose wave
functions $\Psi^{B}$ must satisfy that
\begin{equation}
P^{a_{i} a_{j}}\Psi^{B}\mid _{\xi_{i}\eta_{i} \longleftrightarrow
\xi_{j} \eta_{j}} =\Psi^{B}.
\end{equation}
As $q_{a}$ are (2+1)D Fermi fields, the anticommutators should be used,
and $\Psi^{F}$ must be antisymmetrical under $(a_{i} \longleftrightarrow
a_{j})$ and $((\xi_{i} \eta_{i})
\longleftrightarrow  (\xi_{j} \eta_{i}))$. Namly,
\begin{equation}
P^{a_{i} a_{j}}\Psi^{F}\mid _{\xi_{i}\eta_{i} \longleftrightarrow
\xi_{j} \eta_{j}} =-\Psi^{F}.
\end{equation}
Thus for the 2D Boson case, two solutions of $\Psi^{B}$ can be
constructed as following
\begin{eqnarray}
\Psi^{B}_{1}
= \prod_{i<j} (1-\frac{c}{4}\epsilon (\xi_{ij})\epsilon (\eta_{ij}))
 [ {\cal O_{N}}X^{F}(\xi_{1}\cdots \xi_{N})]
 [ {\cal O_{N}}Y^{F}(\eta_{1}\cdots \eta_{N})], \\
\Psi^{B}_{2}
= \prod_{i<j} (1-\frac{c}{4}\epsilon (\xi_{ij})\epsilon (\eta_{ij}))
 [ {\cal A_{N}}X^{B}(\xi_{1}\cdots \xi_{N})]
  [{\cal A_{N}}Y^{B}(\eta_{1}\cdots \eta_{N})].
\end{eqnarray}
Using eqs.(36),(39),(40) and (42), we can check eq.(45) diractly. In
addition, from the Bethe ansatz equations (25) (26) (30) (31) and $E=
E_{1}+E_{2}$, we can see that the eigenvalues of $\Psi^{B}_{1}$ and
$\Psi^{B}_{2}$ are different each other generally, i.e., the states
corrosponding to $\Psi^{B}_{1}$ and $\Psi^{B}_{2}$ are non-degenerate.

For the 2D Fermion case, the desired results are
\begin{eqnarray}
\Psi^{F}_{1}
= \prod_{i<j} (1-\frac{c}{4}\epsilon (\xi_{ij})\epsilon (\eta_{ij}))
 [ {\cal O_{N}}X^{F}(\xi_{1}\cdots \xi_{N})]
 [ {\cal A_{N}}Y^{B}(\eta_{1}\cdots \eta_{N})], \\
\Psi^{F}_{2}
= \prod_{i<j} (1-\frac{c}{4}\epsilon (\xi_{ij})\epsilon (\eta_{ij}))
 [ {\cal A_{N}}X^{B}(\xi_{1}\cdots \xi_{N})]
  [{\cal O_{N}}Y^{F}(\eta_{1}\cdots \eta_{N})].
\end{eqnarray}
Eq.(46) can also be checked diractly. The eigenvalues corresponding to
$\Psi^{F}$ are also determined by the Bethe equations and
$E=E_{1}+E_{2}$.

 It is similar to ref.\cite{yan} that we can prove $\Psi^{B}_{1,2}$ and
$\Psi^{F}_{1,2}$ shown in above are of
the exact solutions of the eq.(15). Thus
 we conclude that the 2D quantum
many-body problem induced from the quantum DS1 system with
2-component has been solved exactly.

\section{The Ground-State Energies of the System}

In this section, we discuss the ground-state energies of the DS1 system
solved in the previous section by using the Bethe ansatz equations
(25), (26) and (30), (31). Let the 2D N-body problem reduced from 2D DS1
system with 2 colour (or spin) components has $M$ colours down and
$N-M$ colours up. Therefore both $X^{F,B}(\xi_{1},\xi_{2},\cdots \xi_{N})$ and
$Y^{F,B}(\eta_{1}, \eta_{2},\cdots \eta_{N})$ in eqs(47)-(50) are one
dimensional $N-$body wave functions with $M$ colours down and $N-M$
colours up.
We are interested in the limit that $N$, $M$ and the length $L$ of the box
go to infinity proportionately, i.e., both $N/L=D$ and $M/L=D_m$ are finite.

For one dimensional $N$-fermion problem, by the nested Bethe ansatz (or
Bethe-Yang ansatz) equations (25) and (26), the corresponding integration
equations for the ground state read\cite{yang}
\begin{eqnarray}
2\pi\sigma_1&=&-\int_{-B_1}^{B_1}{{2g\sigma_1(\Lambda')d\Lambda'}\over
{g^2+(\Lambda-\Lambda')^2}}
+\int_{-Q_1}^{Q_1}{{4g\rho_1(k)dk}\over
{g^2+4(k-\Lambda)^2}},\\
2\pi\rho_1&=&1
+\int_{-B_1}^{B_1}{{4g\sigma_1(\Lambda)d\Lambda}\over
{g^2+4(k-\Lambda)^2}},
\end{eqnarray}
where $\rho_1(k)$ is particle (i.e.,1D fermion) density distribution
function of $k$, and $\sigma_1(\Lambda)$ is colour-down particle density
distribution function of $\Lambda$. Namely, we have
\begin{eqnarray}
&&D=\int_{-Q_1}^{Q_1}{\rho_1(k)dk},\;\;\;
D_m=\int_{-B_1}^{B_1}{\sigma_1(\Lambda)d\Lambda},\;\;\; \nonumber\\
&&\qquad E_1/N=D^{-1}\int_{-Q_1}^{Q_1}{k^2\rho_1(k)dk}.
\end{eqnarray}

For 1D N-boson case, starting from the nested Bethe ansatz equations (30)
and (31), similar integration equations for ground state of bosons can be
derived (see Appendix). The results are as follows
\begin{eqnarray}
2\pi\sigma_2&=&\int_{-B_2}^{B_2}{{2g\sigma_2(\Lambda')d\Lambda'}\over
{g^2+(\Lambda-\Lambda')^2}}
-\int_{-Q_2}^{Q_2}{{4g\rho_2(k)dk}\over
{g^2+4(k-\Lambda)^2}},\\
2\pi\rho_2&=&1
-\int_{-B_2}^{B_2}{{4g\sigma_2(\Lambda)d\Lambda}\over
{g^2+4(\Lambda-k)^2}}+\int_{-Q_2}^{Q_2}{{2g\rho_2(k')dk'}\over
{g^2+(k-k')^2}},
\end{eqnarray}
where $\rho_2(k)$ and $\sigma_2(\Lambda)$ are bosonic particle density
distribution function of $k$ and its colour-down particle density
distribution function of $\Lambda$ respectively, i.e.,
\begin{eqnarray}
&& D=\int_{-Q_2}^{Q_2}{\rho_2(k)dk},\;\;\;
D_m=\int_{-B_2}^{B_2}{\sigma_2(\Lambda)d\Lambda},\;\;\;\nonumber\\
&&\qquad E_2/N=D^{-1}\int_{-Q_2}^{Q_2}{k^2\rho_2(k)dk}.
\end{eqnarray}

The everage energies of the 2D DS1 ground states described by $\Psi^B_1$,
$\Psi^B_2$, $\Psi^F_1$ and $\Psi^F_2$ (see eqs(47)$-$(50)) are denoted
by $E(\Psi^B_1)$, $E(\Psi^B_2)$, $E(\Psi^F_1)$ and $E(\Psi^F_2)$
respectively. Then, the everage energies per particle for the ground-states
are follows
\begin{eqnarray}
E(\Psi^B_1)/N&=&2E_1/N=2D^{-1}\int_{Q_1}^{Q_1}k^2\rho_1(k)dk, \\
E(\Psi^B_2)/N&=&2E_2/N=2D^{-1}\int_{Q_2}^{Q_2}k^2\rho_2(k)dk, \\
E(\Psi^F_1)/N&=&{1\over N} (E_1+E_2) \nonumber \\
  &=&D^{-1}(\int_{Q_1}^{Q_1}k^2\rho_1(k)dk+\int_{Q_2}^{Q_2}k^2\rho_2(k)dk)
  \nonumber  \\
  &=&{1\over 2}(E(\Psi^B_1)+E(\Psi^B_2)), \\
E(\Psi^F_2)/N&=&E(\Psi^F_1)/N.
\end{eqnarray}
From these equations, the follows can been seen:
1, The everage energies per particle
for the ground states of this 2-dimensional (2D-) DS1
problem are reduced into the everage energies per particle of 1-dimensional
(1D-)many body problems.
As $D$ and $D_m$ are given, by solving the integration
equations $(51)-(56)$, we obtain the $\rho_1(k)$ and $\rho_2(k)$,
and then get the desired results of $E(\Psi^B_1)/N$, $E(\Psi^B_2)/N$,
$E(\Psi^F_1)/N$ and $E(\Psi^F_2)/N$.
2, For the two bosonic
solutions of the 2D DS1 system with 2 colours (eqs (47) (48)), the everage
ground state energies per particle are twice as large as one of 1D-fermions
or 1D-bosons;
3, For the fermion solutions of this 2D-DS1 system, $E(\Psi^F_1)/N$
and $E(\Psi^F_2)/N$ are sum of 1D-fermion everage energy per particle
and 1D-boson's.
4, In general, $E(\Psi^B_1)\neq E(\Psi^B_2)\neq E(\Psi^F_{1,\;{\rm or}\;2})$.
Namely, for same DS1 system, if the statistics of the wave functions
(or particles) is different, the corresponding ground-state energies are
different. This is remarkable and reflects the statistical effects
in the 2D DS1 system.

\section{ Discussions and Summary}

Finally, we would like to speculate some further applications of
the results presented in this paper to the mathematical physics.
Our results may be useful in the following two respects. Firstly,
the Bethe ansatz equations (25), (26) for fermion wave functions
and (30), (31) for boson's can be solved respectively, even though
the equations are systems of transcendental equations for which
the roots are not easy to locate. The so-called string hypothesis
is used for the analysis and classification of the roots for the
Bethe ansatz equations\cite{lai,sch}. Thus, we could study thier
ground state, the excitation and the thermodynamics based on
it\cite{lai,sch}. Then, the thermodynamical properties of the 1D
Bose or Fermi gas with $\delta-$function interaction and with two
components can be explored. The eqs.(47)$-$(50) indicate that
under the thermodynamical limit the 2D DS1 gases (with two color
components) are classified into 2D Bose gases and 2D Fermi gases.
By eqs (47) (48), the 2D Bose gases are composed of two 1D Fermi
gases or 1D Bose gases, and by eqs (49) (50), the 2D Fermi gases
are composed of 1D Fermi gas and 1D Bose gas. Hence, the
thermodynamics of 2D DS1 gases with two color components can be
derived exactly. It would be interesting in physics, because this
is an interesting and nontrivial example to illustrate coupling
(or fusing) of two 1D 2-component gases with $\delta-$function
interaction and with different or same statistics. Secondly, the
colorless DS1 equation originated in studies of nonlinear
phenomena\cite{ds}. Five years ago, Pang, Pu and Zhao\cite{ppz}
showed an example that the solutions of the initial-boundary-value
problem for the related classical DS1 equation in ref.\cite{fokas}
are consistent with the solutions for the quantum DS1 system with
time-dependent applied forces. This indicates that the classical
solutions of DS1 equation are corresponding to the classical limit
of the solutions for the quantum DS1 system. This is actually a
new method to reveal the solutions of the colorless DS1 equation.
To the quantum DS1 system with color indices studied in this
present paper, similar correspondences are expectable. Hence, the
structure of the solutions of the quantum DS1-system with color
indices revealed in this paper would be helpful to understand the
corresponding classical solutions of DS1 systems with color. The
specific studies on the above speculations would be meaningful,
however, they are beyond the scope of this present paper.

To summarize. We formulated the quantum multi-component DS1 system
in terms of the quantum multi-component many-body Hamiltonain in
2D space. Then we reduced this 2D Hamiltonain to two 1D
multi-component many-body problems. As the potential between two
particles with two components in one dimension is
$\delta-$function, the Bethe-Yang ansatz was used to solve these
1D problems. By using the ansatz of ref.\cite{pang} and
introducing some useful Young operators, we presented a new N-body
variable-separation ansatz for fusing two 1D-solutions to
construct 2D wave functions of the quantum many-body problem which
is induced from the quantum 2-component DS1 system. There are two
types of wave functions: Boson's and Fermion's. Both of them
satisfy the 2D many-body Schr\"{o}dinger equation of the DS1
system exactly. The results have been used to study the ground
states of the system. Some further applications of the results
presented in this paper are speculated and discussed. This paper
serves as a review to refs.\cite{pang,yan,yan1}.

\section*{Appendix}
\setcounter{equation}{0}
\renewcommand{\theequation}{A.\arabic{equation}}

Let us derive eqs (54) and (55) in the text. We start from the
Bethe ansatz equations (30) and (31) of 1D bosons with two color
components. Taking the logarithm of (30) and (31) respectively, we
have
\begin{eqnarray}
&&k_jL=2\pi I_k-2\sum\limits_{i=1}^{N}\tan^{-1}{{k_j-k_i}\over{g}}
      -2\sum\limits_{\beta=1}^{M}\tan^{-1}{{2(\Lambda_\beta-k_j)}\over{g}}\\
&&2\sum\limits_{\alpha=1}^{M}\tan^{-1}{{\Lambda_\beta-\Lambda_\alpha}
\over{g}}=2\pi J_\Lambda
+2\sum\limits_{j=1}^{N}\tan^{-1}{{2(\Lambda_\beta-k_j)}\over{g}},
\end{eqnarray}
where (for the case of $N=$even, $M=$odd)
\begin{eqnarray*}
{{1}\over{2}}+I_k&=&{\rm successive\;integers\;from}\;1-{1\over2}N\;\;
{\rm to}\;+{1\over2}N, \\
J_\Lambda&=&{\rm successive\;integers\;from}\;-{1\over2}(M-1)\;{\rm to}\;
+{1\over2}(M-1).
\end{eqnarray*}
We can now approach the limit $N\longrightarrow \infty$,
$M\longrightarrow \infty$, $L\longrightarrow \infty$ proportionally, obtaining
\begin{eqnarray}
&&k=2\pi f_2-2\int_{-Q_2}^{Q_2}dk'\rho_2(k')\tan^{-1}{{(k-k')}\over{g}} \nonumber\\
&&\qquad -2\int_{-B_2}^{B_2}d\Lambda\sigma_2(\Lambda)\tan^{-1}{{2(\Lambda-k)}\over{g}},\\
&&2\int_{-B_2}^{B_2}d\Lambda'\sigma_2(\Lambda')
\tan^{-1}{{\Lambda-\Lambda'}\over{g}}=2\pi h_2 \nonumber\\
&&\qquad +2\int_{-Q_2}^{Q_2}dk\rho_2(k)\tan^{-1}{{2(\Lambda-k)}\over{g}},\\
&&{{dh_2}\over{d\Lambda}}=\sigma_2,\qquad \qquad {{df_2}\over{dk}}=\rho_2,\\
&&D={N\over L}=\int_{-Q_2}^{Q_2}{\rho_2(k)dk},\qquad
D_m={M\over L}=\int_{-B_2}^{B_2}{\sigma_2(\Lambda)d\Lambda}.
\end{eqnarray}
Or, after differentiation,
\begin{eqnarray}
2\pi\sigma_2&=&\int_{-B_2}^{B_2}{{2g\sigma_2(\Lambda')d\Lambda'}\over
{g^2+(\Lambda-\Lambda')^2}}
-\int_{-Q_2}^{Q_2}{{4g\rho_2(k)dk}\over
{g^2+4(k-\Lambda)^2}},\\
2\pi\rho_2&=&1
-\int_{-B_2}^{B_2}{{4g\sigma_2(\Lambda)d\Lambda}\over
{g^2+4(\Lambda-k)^2}}+\int_{-Q_2}^{Q_2}{{2g\rho_2(k')dk'}\over
{g^2+(k-k')^2}},
\end{eqnarray}
which are just eqs (54) and (55).

\end{document}